\documentclass[amssymb, reprint, showpacs,  longbibliography]{revtex4-1}				  

\usepackage{graphicx}
\usepackage{dcolumn}
\usepackage{amsthm} 
\usepackage{amsmath,amsfonts,amssymb}
\usepackage{bm}
\usepackage{extarrows}  
\usepackage{chemarrow}  
\usepackage{booktabs} 
\usepackage[utf8]{inputenc}
\usepackage[T1]{fontenc}
\usepackage[overload]{empheq}

\usepackage{mathptmx} 

\begin{document}

\title{Restrictions on wave equations for passive media}

\author{Sverre \surname{Holm}} \affiliation{Department of Informatics, University of Oslo, P.~O.~Box 1080, N--0316 Oslo, Norway}

\author{Martin \surname{Blomhoff Holm}} \affiliation{Department of Economics, BI Norwegian Business School, N--0442 Oslo, Norway}

\date{\today}

\begin{abstract}
Most derivations of acoustic wave equations involve ensuring that causality is satisfied. Here we explore the consequences of also requiring that the medium should be passive.  This is a stricter criterion than causality for a linear system and implies that there are restrictions on the relaxation modulus and its first few derivatives.  The viscous and relaxation models of acoustics satisfy passivity and have restrictions on not only a few, but all derivatives of the relaxation modulus. This is the important class of completely monotone systems. It is the only class where the medium is modeled as a system of springs and dampers with positive parameters. It is shown here that the attenuation as a function of frequency for such media has to increase slower than a linear function. Likewise the phase velocity has to increase monotonically. This gives criteria on which one may judge whether a proposed wave equation is passive or not, as illustrated by comparing two different versions of the viscous wave equation.
\end{abstract}

							 %
\maketitle


\section{Introduction}

In 1981 Weaver and Pao \cite{Weaver1981} showed that a wave equation's asymptotic value for attenuation has to  increase slower than a linear function of frequency. They claimed that this result followed from causality, passivity and linearity. This result seems to have gone unnoticed among researchers in acoustics. The usual criterion to apply to a wave equation's solution is causality, assume that linearity holds, and to overlook the passivity requirement.  
A recent example \cite{buckingham2015frequency} shows that the asymptote of the attenuation may increase with any power. This seems to contradict Weaver and Pao's result, but since Ref.~\citenum{buckingham2015frequency} only takes causality into account, the criterion is weaker than in Ref.~\citenum{Weaver1981}. 


One reason why Weaver and Pao's result and the more recent work on this by Hanyga \cite{seredynska2010relaxationdispersion, Hanyga2013Wave,  hanyga2014dispersion} may have been overlooked is that the results are quite abstract and mathematical. Further, some of the work above also assumes that the material can be modeled with a spring damper model without much argument for why this is so. This paper is an attempt to justify this assumption and present the result in a more accessible way.


To illustrate the point, we use the viscous wave equation.
It usually comes in two versions, with either of the loss terms below. A natural question to ask is if they are completely equivalent or if one or the other has properties which the other one does not possess. 
\begin{equation}
\nabla^2 u  - \frac{1}{c_0^2} \frac{\partial^2 u}{\partial t^2} + 
\begin{cases}
	\frac{\tau}{c_0^2} \frac{\partial^3 u}{\partial t^3} = 0\\
	\tau \frac{\partial}{\partial t} \nabla^2 u = 0
\end{cases}
\label{eq:viscousX2}
\end{equation}
Here $u$ can be particle displacement or pressure, $c_0$ is the speed of sound as frequency approaches zero frequency, and $\tau$ is the ratio of a viscosity and an elastic modulus. The upper version is often seen in the literature on nonlinear acoustics \cite{Hamilton98}, while the lower version has a history that goes back to Stokes in 1845. The reference is p.~302 of Ref.~\citenum{Stokes1845}. 

Note that the attenuation is the same in both cases if we assume low losses or low frequencies.
This can be seen from the fact that in that case one can convert one equation to the other by assuming first that the term in the curly brackets can be neglected so that
%
$\nabla^2 u \approx \frac{1}{c_0^2}\frac{\partial^2 u}{\partial t^2}$,
and then replacing the Laplacian with the temporal derivative or vice versa in the loss term. 

\begin{figure}[tb]
\centering
\includegraphics[width=0.7\columnwidth]{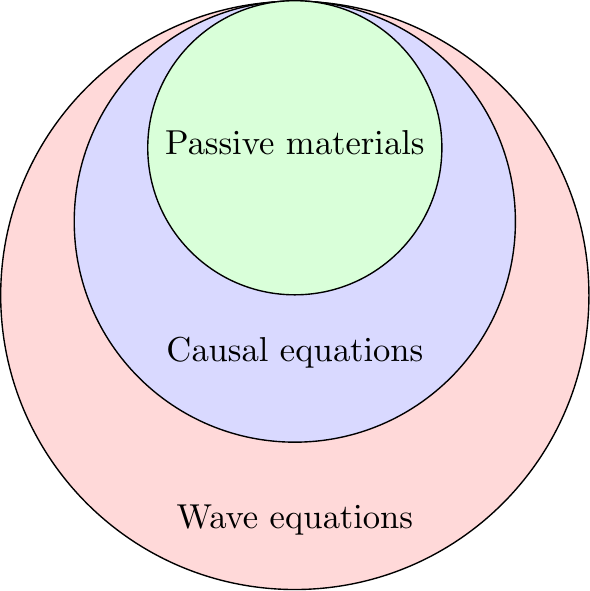}
 \caption{The set of passive, linear materials is a subset of causal wave equations, which again is a subset of possible wave equations (color online)}
  \label{fig:Passive3}
\end{figure}

The claim in the present paper is that the set of causal wave equations is larger than the set of passive, linear materials with wave equations. Furthermore, the set of possible wave equations is even larger. This is illustrated in Fig.~\ref{fig:Passive3}. 
%

We start by analyzing three different forms of the constitutive equation. Then it is shown that a network of springs and dampers, where each branch has positive elasticities and viscosities, guarantees that a system is passive. This result implies that the relaxation modulus possesses a property called complete monotonicity.
A consequence of this property is that the complex wavenumber is a complete Bernstein function. Such functions have specific  properties that imply that the asymptotic properties of the attenuation and dispersion can be found. This is applied to the two viscous wave equations of \eqref{eq:viscousX2} in order to judge which one that corresponds to a physically realizable passive medium.

\section{Constitutive equations}
In acoustics, one form of a linear constitutive relation between pressure variations, $p$, around an equilibrium pressure, $p_0$, and relative density variations, $\rho/\rho_0$, is:
\begin{align}
p(t) =  h(t)* \frac{\rho(t)}{\rho_0}.
\label{eq:convH(t)Physical}
\end{align}
In linear viscoelasticity, where much of the theory of this paper is taken from, the impulse response, $h(t)$, is hard to measure due to the impracticality of impulsive sources, so the step response, $G(t)$, where $h(t) = dG(t)/dt$, is used instead. Also pressure is replaced by strain, $\sigma(t)=-p(t)$, and density by stress, $\epsilon(t) = -\rho(t)/\rho_0$. The 
1-D constitutive equation of a linear system, from Ch.~2 of Ref.~\citenum{Mainardi2010}, is then:
\begin{align}
\sigma(t) =  G(t)* \frac{\partial \epsilon(t)}{\partial t}.
\label{eq:convConstPhysical}
\end{align}
The kernel $G(t)$ is called the relaxation modulus. A linear viscoelastic material can be described in three different ways and they will be related to each other here.

\subsection{Linear differential equation model}
The first description is a linear differential equation between stress and strain with constant coefficients:
\begin{align}
	\left[1 + \sum_{k=1}^{p} a_k \frac{\partial^{k}}{\partial t^{k}}\right] \sigma(t)  = \left[E_e + \sum_{k=1}^{q} b_k \frac{\partial^{k}}{\partial t^{k}}\right] \epsilon(t).
\label{eq:highOrder}
\end{align}
Ch.~2.1 of Ref.~\citenum{tschoegl1989phenomenological} and many other sources state that the differential equation description is equivalent to the superposition model of \eqref{eq:convConstPhysical}.

\subsection{The causal fading memory  model}
The second description is the convolution model of \eqref{eq:convConstPhysical} where there are restrictions on the kernel, $G(t)$,  in order for it to represent a fading memory. Then changes in the past have less effect now than more recent changes. This is the hereditary model of Boltzmann \cite{boltzmann1876theorie,markovitz1977boltzmann}. To ensure causality, the kernel of  \eqref{eq:convConstPhysical} also has to be zero for negative time.

The fading memory concept is not always well defined. It certainly implies that the kernel has to be non-negative and that it is monotonically decreasing. It may also imply conditions on the derivatives of  higher orders. In order to reach a strict definition of fading, the concept of passivity needs to be introduced.

\subsubsection{The passive fading memory model}
There are at least two tests for passivity in the literature. The first is the requirement that the dissipation is non-negative at all time \citep{akyildiz1990spring}
\begin{equation}
D(t)=\int_{0}^t \sigma(\tau) \frac{\partial \epsilon(\tau)}{\partial \tau} d\tau \ge 0.
\label{eq:Dissipation}
\end{equation}
To ensure that no energy is produced by the system prior to application of the input, it is necessary that the system is causal \cite{triverio2007stability}. Therefore the set of passive systems is a true subset of the causal system group in Fig.~\ref{fig:Passive3}, and the lower limit in the integral  can be 0 rather than $-\infty$. This also implies that there is no need to test for causality with e.g.~the Kramers-Kronig relation \cite{waters2005causality} when passivity is satisfied.

The second measure of passivity is the more general Clausius-Duhem inequality which is an expression of the second law of thermodynamics or non-negativity of the rate of entropy production  \citep{lion1997thermodynamics}  
\begin{equation}
\rho \psi(t) + D(t) \ge 0,
\label{eq:Clausius-Duhem}
\end{equation}
where 
$\psi(t)$ with $\psi(0)=0$, is the isothermal free energy which equals the total strain energy stored in
the springs.



In Ref.~\citenum{akyildiz1990spring} a fading memory model which had a positive monotonously falling relaxation modulus with no requirements on the higher order derivatives, was found to  give a non-negative dissipation according to \eqref{eq:Dissipation}. 
Here we will adopt the stricter result of Ref.~\citenum{haupt2002finite} where it is shown that the memory kernel also has to be convex  for it to satisfy the  requirement of a non-negative rate of entropy production. 

Thus the passive fading memory condition adopted here is:
\begin{equation}
(-1)^n \frac{d^n G(t)}{dt^n} \ge 0, \quad t>0, \quad n=0, 1, 2
\label{eq:Fading}
\end{equation}
i.e. the kernel, $G(t)$, is non-negative, non-increasing, and convex.



\subsection{Spring-damper model and complete monotonicity}

The third description is the spring-damper model which we introduce with two examples.

%
%
\subsubsection{Kelvin-Voigt model}
\begin{figure}[tb]
\centering
\includegraphics[width=0.8\columnwidth]{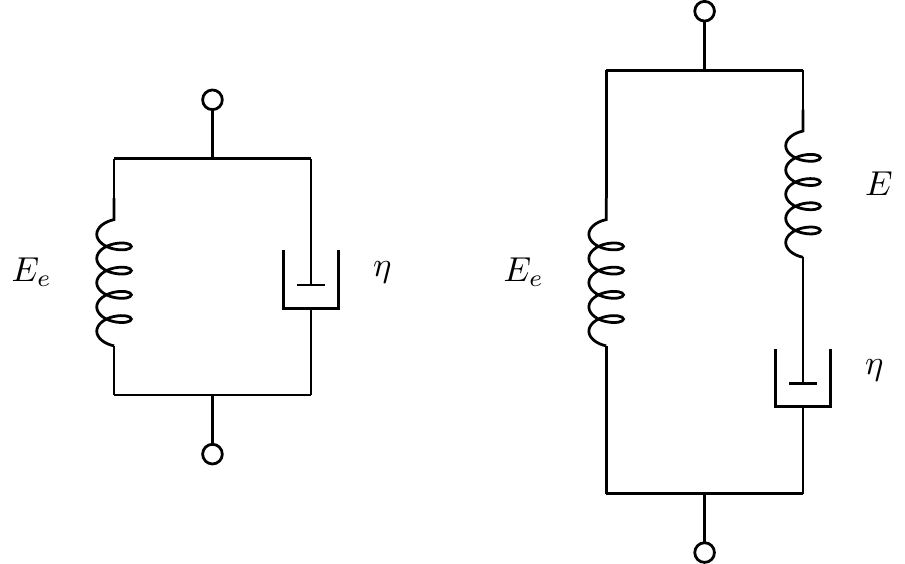}
  \caption{Kelvin-Voigt and Zener (right) constitutive models}
  \label{fig:Kelvin-Voigt_Zener}
\end{figure}

\begin{figure}[tb]
\centering
\includegraphics[width=0.7\columnwidth]{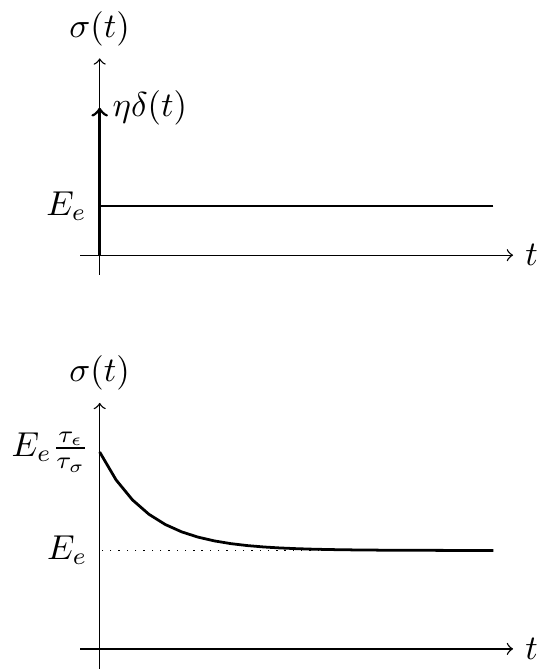}
  \caption{Relaxation moduli of Kelvin-Voigt and Zener (lower) constitutive models}
  \label{fig:Kelvin-Voigt_ZenerRelaxation}
\end{figure}

The left-hand model in Fig.~\ref{fig:Kelvin-Voigt_Zener} is the Kelvin-Voigt model of linear viscoelastivity. The model is expressed by a first order differential operator on the right-hand side of \eqref{eq:highOrder}, i.e. $p=0$, $q=1$, $a_1=0$, and  $b_1=\eta$ which is the viscosity. 
The relaxation modulus, the stress response to a unit step in strain, for the Kelvin-Voigt model is:
\begin{equation}
G(t) = E_e  + \eta  \delta(t)
\label{eq:KV-Relaxation}
\end{equation}
and is plotted in the upper part of Fig.~\ref{fig:Kelvin-Voigt_ZenerRelaxation}. Observe that it is causal and fading in the sense of \eqref{eq:Fading}.
%

The Kelvin-Voigt model leads to the viscous wave equation. It is  common in acoustics instead to derive 
the viscous wave equation from the Navier-Stokes equation combined with conservation of mass and an equation of state which only describes elasticity \cite{blackstock2000fundamentals}. In that case, the material property of viscosity is mixed with the conservation of linear momentum, making it harder to distinguish empirical material properties from fundamental conservation laws. Conservation of momentum and energy (equivalent to conservation of mass in the non-relativistic case and thus the continuity equation) express fundamental physical principles in the form of invariance to spatial and temporal translations as stated in Noether's theorem, see e.g.~Ch.~II of Ref.~\citenum{landau1976mechanics}. 
Whether all the material properties are combined in the constitutive equation as in the Kelvin-Voigt model or not, the end result is the same so these are just alternative ways of deriving Stokes' viscous wave equation.
\subsubsection{Zener model}
In the Zener model, the right-hand model in  Fig.~\ref{fig:Kelvin-Voigt_Zener}, a first order differential term is added on the left-hand side of \eqref{eq:highOrder} so $p=q=1$, $a_1= \tau_\sigma$, $b_1=\eta$. The time constants in terms of the physical components are:
\begin{equation}
\tau_\sigma = \eta/E_1 \le  \tau_\epsilon = \eta/E', \quad \frac{1}{E'} = \frac{1}{E_e} + \frac{1}{E}.
\label{eq:ZenerCondition}
\end{equation}
The Zener model's relaxation modulus is:
\begin{equation}
G(t) = E_e + E_e(\frac{\tau_\epsilon}{\tau_\sigma}-1) e^{-t/\tau_\sigma}.
\label{eq:Zener-Relaxation}
\end{equation}
It is plotted in the lower plot of Fig.~\ref{fig:Kelvin-Voigt_ZenerRelaxation} and it is causal and convex. 
The Zener model is in fact the equation of state in the relaxation model in acoustics although that is not how it is commonly presented. An example is 
Ref.~\citenum{blackstock2000fundamentals} which postulates an equation of state between pressure and density variations:
\begin{equation}
\tau \left( \frac{\partial p}{\partial t} -c_\infty^2 \frac{\partial \rho}{\partial t}\right) + (p - c_0^2 \rho ) = 0,
\label{eq:BlackstockRelax}
\end{equation}
where $c_0$ and $c_\infty$ are the asymptotic values for phase velocity for low and high frequency.
%

This  is in fact the Zener constitutive equation  with $p=-\sigma$ and $\rho/\rho_0=-\epsilon$, although it is not  identified as such.

If the two time constants are very near each other, $\tau_\sigma  \lesssim \tau_\epsilon$, or the equivalent  $c_0  \lesssim c_\infty$, this model describes a relaxation model as briefly discussed in Ref.~\citenum{holm2011}, and it is a building block in the multiple relaxation models for salt water and air.

\subsubsection{The relaxation spectrum}

\begin{figure}[tb]
\centering
\includegraphics[width=0.8\columnwidth]{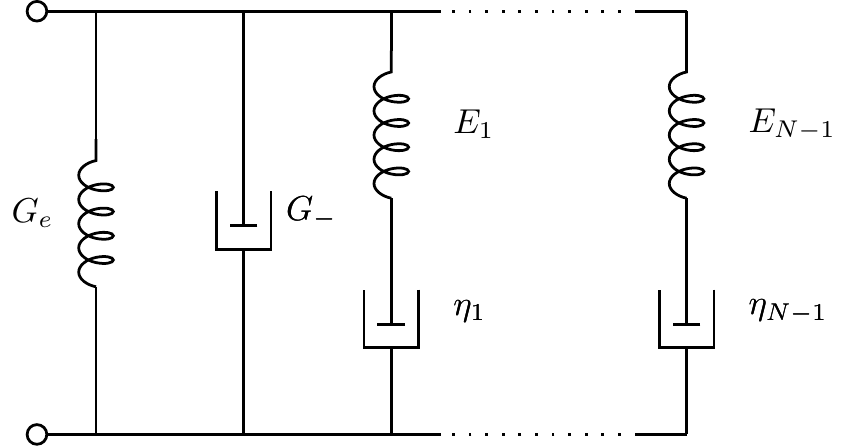}
  \caption{The  model for a general viscoelastic material}
  \label{fig:Maxwell-Wiechert-damper}
\end{figure}

The examples demonstrate that both relaxation moduli are causal and fading. They also show that the relaxation modulus may consist of a constant, a weighted  impulse at time zero, and a weighted sum of exponentials as in Eq.~(2.28) of Ref.~\citenum{Mainardi2010}. The latter is a  series with real, positive coefficients:
\begin{equation}
G(t) = G_e + G_-\delta(t) + G_\tau(t),
\enspace G_\tau(t) = \sum_{n=1}^{N} E_n \exp(-t/\tau_n),
\label{eq:RelaxationSpectrum}
\end{equation}
where the terms in the sum represent series combination of springs and dampers in parallel. They may be in parallel with a spring $G_e\ge 0$ and another parallel damper, $G_- \ge 0$ as shown in Fig.~\ref{fig:Maxwell-Wiechert-damper}.
%
A realistic model is obtained if $G_\tau(t)$ is modeled 
with spring constants, $E_n$ and viscosities, $\eta_n = \tau_n E_n$, which are non-negative. The series expansion for $G_\tau(t)$  is a Prony series or a general Dirichlet series with positive weights.

\subsubsection{Complete monotonicity}
\label{sec:CM}

The relaxation spectrum of \eqref{eq:RelaxationSpectrum} satisfies a general pattern where the derivatives switch signs or are zero for all orders:
\begin{equation}
(-1)^n \frac{d^n G(t)}{dt^n} \ge 0, \quad t>0, \quad n=0, 1, 2, \ldots
\label{eq:CM}
\end{equation}
The criterion is stricter than that of \eqref{eq:Fading} and is called complete monotonicity.
%
%
The limit of the sum in \eqref{eq:RelaxationSpectrum} is a Laplace transform:
\begin{equation}
G(t) = \int_0^\infty G(s) e^{-st} ds.
\label{eq:LaplaceRelaxation}
\end{equation} 
Any model that satisfies complete monotonicity can be expressed as a Laplace transform, and any complete monotonous function has a non-negative Laplace transform (Ref.~ \citenum{schilling2012bernstein}, definition 1.3).

An exponential solution, $e^{-t/\tau}$, such as in  \eqref{eq:RelaxationSpectrum}, is one of the simplest examples of a response function where the derivative switches sign according to \eqref{eq:CM}. 
There are also fractional generalizations of the Kelvin-Voigt and Zener constitutive equations where the first-order derivative is replaced by a non-integer derivative of order $\alpha$. At first sight, these models may look very different from spring-damper models. The relaxation moduli of these models are described either by power law functions,  $t^{\alpha}, \alpha\le 0$, or by Mittag--Leffler function. But both of these functions are also completely monotone \cite{Mainardi2010}. Actually, these relaxation moduli can also be described by an infinite relaxation series of the form of \eqref{eq:RelaxationSpectrum} as shown in Refs.~\citenum{gross1947creep,caputo1971linear, holm2017spring}.

\section{The special role of spring damper models}
\begin{figure}[tb]
\centering
\includegraphics[width=0.7\columnwidth]{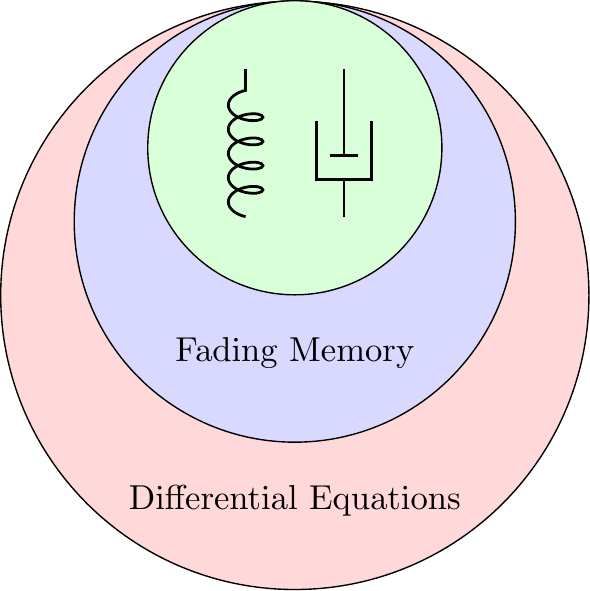}
 \caption{The set of spring-damper models with physical parameters is a subset of fading memory models, which itself is a subset of all solutions to linear differential equations (color online)}
  \label{fig:FadingModels}
\end{figure}

The relationship between the three descriptions, the linear differential equation, the fading memory,  and the spring-damper network is shown in Fig.~\ref{fig:FadingModels}. The solution to a linear differential equation can always be written in the form of convolution integral of the form of \eqref{eq:convConstPhysical}. But only a subset of these solutions satisfy the fading memory criterion. The fading memory subset further contains two cases:
\begin{enumerate}
\item The  relaxation spectrum of a physical model of form \eqref{eq:RelaxationSpectrum} can always be written as a linear differential equation of the form of \eqref{eq:highOrder} and it will always produce the fading memory kernel of \eqref{eq:convConstPhysical}. Thus the subset of physical models exists as a subset of the two other descriptions.
\item There exist fading memory models which do not correspond to a spring damper system with positive coefficients.
\end{enumerate}
\subsection{A spring damper system with non-physical parameters}
To illustrate the last point, we analyze the previously mentioned  example of a model with a non-negative dissipation \cite{akyildiz1990spring}. This model has a positive monotonously falling relaxation modulus with no requirements on the higher order derivatives, and it was formed by a combination of a physical spring-damper and one with non-physical parameters:
\begin{equation}
 G(t) = E_1 \exp(-t/\tau_1)  + E_2 \exp(-t/\tau_2), \quad E_2 <0.
\label{eq:non-physical}
\end{equation}
For illustration, we assume that the first time-constant is greater than the second one, $\tau_1 > \tau_2 > 0$. Because the time constant is the ratio of viscosity and elasticity, a non-physical negative elasticity, $E_2$, implies  that the second viscosity also is non-physical and negative, $\eta_2 <0$. 
%
%
In Ref.~\citenum{akyildiz1990spring} it was found that the conditions for $G(0) \ge 0$ and $G'(0) \le 0$ are that $E_1+E_2 \ge 0$ and $E_1/\tau_1 + E_2/\tau_2 \ge 0$. The first condition implies that the spring constant of the physical spring is greater than that of the negative spring, and the second condition also implies that the total dissipation, \eqref{eq:Dissipation}, is  non-negative. The interpretation is that the energy produced by the last spring and damper is more than absorbed by the first one.

We are interested in a condition on the second derivative in order to have a non-negative rate of entropy, \eqref{eq:Clausius-Duhem}. By differentiating and setting the derivative at $t=0$ of order $n$ to zero, the condition of an alternating sign of the derivative, as in \eqref{eq:CM}, is satisfied up to and including order $n$ if the following condition holds:
\begin{equation}
\frac{E_1}{\tau_1^n} + \frac{E_2}{\tau_2^n} \ge 0.
\label{eq:Condition}
\end{equation}
In the limit as $n \rightarrow \infty$, the second elasticity, $E_2$, has to approach zero in order to satisfy this requirement because $\tau_1$ is the greatest of the two time constants. 

When $n=2$ in \eqref{eq:Condition}, this is an example of a fading memory system which satisfies the requirement for non-negative rate of entropy production. It is a system consisting of a physical branch and a non-physical branch. We may therefore dismiss many of the fading memory systems that satisfy the Clausius--Duhem inequality if we impose the additional criterion that each subbranch should be a physical system, i.e. have positive elasticity and viscosity.


\subsection{Complete monotone models}

As shown in the previous section, the two main models of acoustics are spring-damper models. Also in linear viscoelasticity, the constitutive equations in the form of spring-damper models play a fundamental role \cite{tschoegl1989phenomenological}.

Here we therefore surmise that completely monotone models, i.e. those based on springs and dampers, play a special role in fading memory systems. The hypothesis is that they are better physically motivated than models which only have positive, monotone, and convex relaxation moduli, despite that these models also satisfy the Clausius-Duhem inequality. The spring-damper models may however only be a subset of the interesting models, but a very important subset.

%


\section{Analysis of the wave equation}
The fact that most of the interesting systems have a relaxation modulus, $G(t)$, which is a completely monotone function has consequences for the asymptotic properties of the solution to the wave equation.

\subsection{Wavenumber as a function of relaxation modulus}
First we need to express the wave equation in terms of the relaxation function. The following  is a less rigorous version of the wave equation derivation of Ref.~\citenum{hanyga2014dispersion}.
It builds on transforming the main equations to the frequency domain \cite{holm2011}. The constitutive equation \eqref{eq:convConstPhysical} is then 
\begin{equation}
\sigma(\omega) 
= i\omega G(\omega) \epsilon(\omega), 
\label{eq:Constitutive}
\end{equation}

The frequency domain version of the  conservation of momentum is:
\begin{align}
 \rho_0 \frac{\partial^2 u}{\partial t^2} =  \nabla \mathbf{\sigma} \quad \Leftrightarrow \quad 
	\rho_0 (i \omega)^2 u(\omega) = -i k\: \sigma(\omega) .
	\label{Eq:ConservationMomentumFrequencyPhysical}
\end{align}
and conservation of mass gives:
\begin{align}
\epsilon(t) = \frac{\partial u}{\partial x} \quad  \Leftrightarrow \quad
	\epsilon(\omega) = -i k\:u(\omega).
	\label{Eq:ConservationMassFrequencyPhysical}
\end{align}
Now insert \eqref{Eq:ConservationMassFrequencyPhysical} in \eqref{eq:Constitutive} to eliminate $\epsilon(\omega)$ and then use \eqref{Eq:ConservationMomentumFrequencyPhysical} to eliminate $u(\omega)$. The wave number is then:
\begin{equation}
k^2(\omega)= \frac{ \rho_0 \omega^2}{ i\omega G(\omega)}. 
\label{eq:dispersionPhysical}
\end{equation}
%
Equation \eqref{eq:dispersionPhysical} is now made more general by  using the Laplace transform  rather than the Fourier transform, i.e.~substitution of $s=i \omega$:
\begin{equation}
K^2(s) =-k^2(s) =  \rho_0 s^2 \frac{1}{ s G(s)}.
\end{equation}
The new variable $K$ is related to the wavenumber by $K=i k$. Taking the square root gives:
\begin{equation}
K(s) = ik(s) =  \sqrt{\rho_0}  \frac{s}{\sqrt{s G(s)}}.
\label{eq:Hanyga2014Eq15}
\end{equation}
This shows how the wavenumber depends on the relaxation modulus and  corresponds to Eq.~(15) in Ref.~\citenum{hanyga2014dispersion}.

\subsection{Bernstein property}
\label{sec:BernsteinProperty}
In order to proceed, we will need the definition of a  class of functions which are  related to completely monotone ones. Bernstein functions are non-negative functions where the derivative is completely monotone:
\begin{equation}
f(t) \ge 0, \quad (-1)^{n} \frac{d^n f(t)}{dt^n} \le 0, \quad t>0, \quad n=1, 2, \cdots
\label{eq:BernsteinDerivatives}
\end{equation}
Examples of such functions are $1-e^{-t/\tau}$ and $t^{\alpha}, 0\le \alpha \le 1$.

A subclass of Bernstein functions is the set of complete Bernstein functions. 
%
These functions are defined in Appendix \ref{app:cbf} and will be called CBF from now on. Likewise completely monotone functions will be denoted by CM.

If the relaxation modulus $G(t)$ is CM, then the wavenumber $K(s)$ is a CBF. The argument is formally given in Appendix \ref{app:Proof} and in Ref.~\citenum{seredynska2010relaxationdispersion}, and follows these steps:
\begin{enumerate}
\item  If $G(t)$ is CM, then the Laplace transform $G(s)$  as given by \eqref{eq:LaplaceRelaxation} is also CM, since it is a combination of a positive constant, a positive impulse, and positively weighted exponentials
\item If $G(s)$ is CM, then $sG(s)$ is CBF
\item A CBF  applied to  a second CBF produces a third CBF. Both $sG(s)$ and the square root are CBF, so then $\sqrt{sG(s)}$ is also CBF
\item If $\sqrt{sG(s)}$  is CBF then $\frac{s}{\sqrt{s G(s)}}$ is CBF
\end{enumerate}
The consequence of this argument is that the wavenumber $K(s)$ of \eqref{eq:Hanyga2014Eq15}  is a complete Bernstein function.

\subsection{Consequences of the Bernstein property}

If $K(s)$ is a CBF, the same applies to $k(s) = K(s)/i$. Every CBF can be written in the following form (exact definition in Appendix \ref{app:cbf}):
\begin{equation}
K(s) = a + b s + \beta(s),
\label{eq:BernsteinDecomp1}
\end{equation}
where $a, b \ge 0$, $a = K(0)$, $b=\lim_{s\to\infty} K(s)/s$ and $\beta(s)$ is a term with sublinear growth. 
That means that the exponent $y$ in a power law expression $s^y$ has to be less than or equal to one as $s \rightarrow \infty$.
%
%
%

Often this expression can be even more simplified by noting that the constant $a=K(0)$ may be zero. From \eqref{eq:Hanyga2014Eq15} that means that one needs to find $\lim_{s\to 0} sG(s) = \lim_{t\to \infty} G(t)$. The last expression is found from the final value theorem for the Laplace transform. In Ref.~\citenum{Hanyga2013Wave} it is argued that this limit always is positive, but from the expression for the relaxation modulus in \eqref{eq:RelaxationSpectrum} it can be seen that the limiting value is $G_e$, the spring across the terminals in Fig.~\ref{fig:Maxwell-Wiechert-damper}. The presence of this spring is what distinguishes a solid from a liquid \cite{tschoegl1989phenomenological}, so only for those materials that can be considered to be solids one has  $\lim_{s\to 0} sG(s) > 0$ and thus $a=0$. In contrast to Ref.~\citenum{Hanyga2013Wave}, the following argument is however not dependent on $a$ being 0.

The wave number can also be expressed by the  attenuation, $\alpha(\omega)$, and the  phase velocity, $c_p(\omega)$;
\begin{equation}
k (\omega) =  \frac{\omega}{c_p(\omega)} - i \alpha(\omega) = \omega \left(\frac{1}{c_\infty} + \frac{d(\omega)}{\omega} \right)- i \alpha(\omega),
\label{eq:knumberDecomposed}
\end{equation}
The phase velocity has been decomposed into a constant asymptotic value and a frequency dependent component $d(\omega)$ which is the excess dispersion. 
Expressed with Laplace transforms this is
\begin{equation}
K(s) = ik(s) =  \frac{s}{c_\infty} + id(s) +  \alpha(s).
\label{eq:KDecomposed}
\end{equation}
This expression is combined with \eqref{eq:BernsteinDecomp1}, noting that $b=1/c_\infty$. This parameter may be 0 for instance for the wave equation derived from the Kelvin-Voigt equation where $c_\infty=\infty$ \cite{Holm2010}. Equating real and imaginary parts gives
\begin{equation}
\alpha(s) = \Re \beta(s) + a, \quad d(s) = \Im \beta(s).
\label{eq:BernsteinK2}
\end{equation}

\subsection{Asymptotic properties}

Equation \eqref{eq:BernsteinK2} shows that both the attenuation and the excess dispersion are proportional to $\beta(s)$, which is a term with sublinear growth. 
This implies that both attenuation $\alpha(\omega)$ and  excess dispersion $d(\omega)$ also must have sublinear growth. For the attenuation this means:  
\begin{equation}
\lim_{\omega\to\infty} \alpha(\omega) / \omega =  0 \quad \textbf{or} \quad \lim_{\omega\to\infty} \alpha(\omega) \propto \omega^y, \enspace y\le 1.
\label{eq:asymptoteAttenuation}
\end{equation}
The asymptotic result for the excess dispersion compared to \eqref{eq:knumberDecomposed} implies that the phase velocity is a non-decreasing function of frequency:
\begin{equation}
c_p(\omega) \ge 0, \quad \frac{d c_p(\omega)}{d \omega)} \ge 0.
 \label{eq:asympoteDispersion}
\end{equation}



%
%
%
%


\section{Viscous wave equations}
The results of the previous section  are the tools required to decide among the two viscous wave equations of \eqref{eq:viscousX2}.

\subsection{Temporal derivative in loss term}

\begin{figure}[tb]
\centering
\includegraphics[width=\columnwidth]{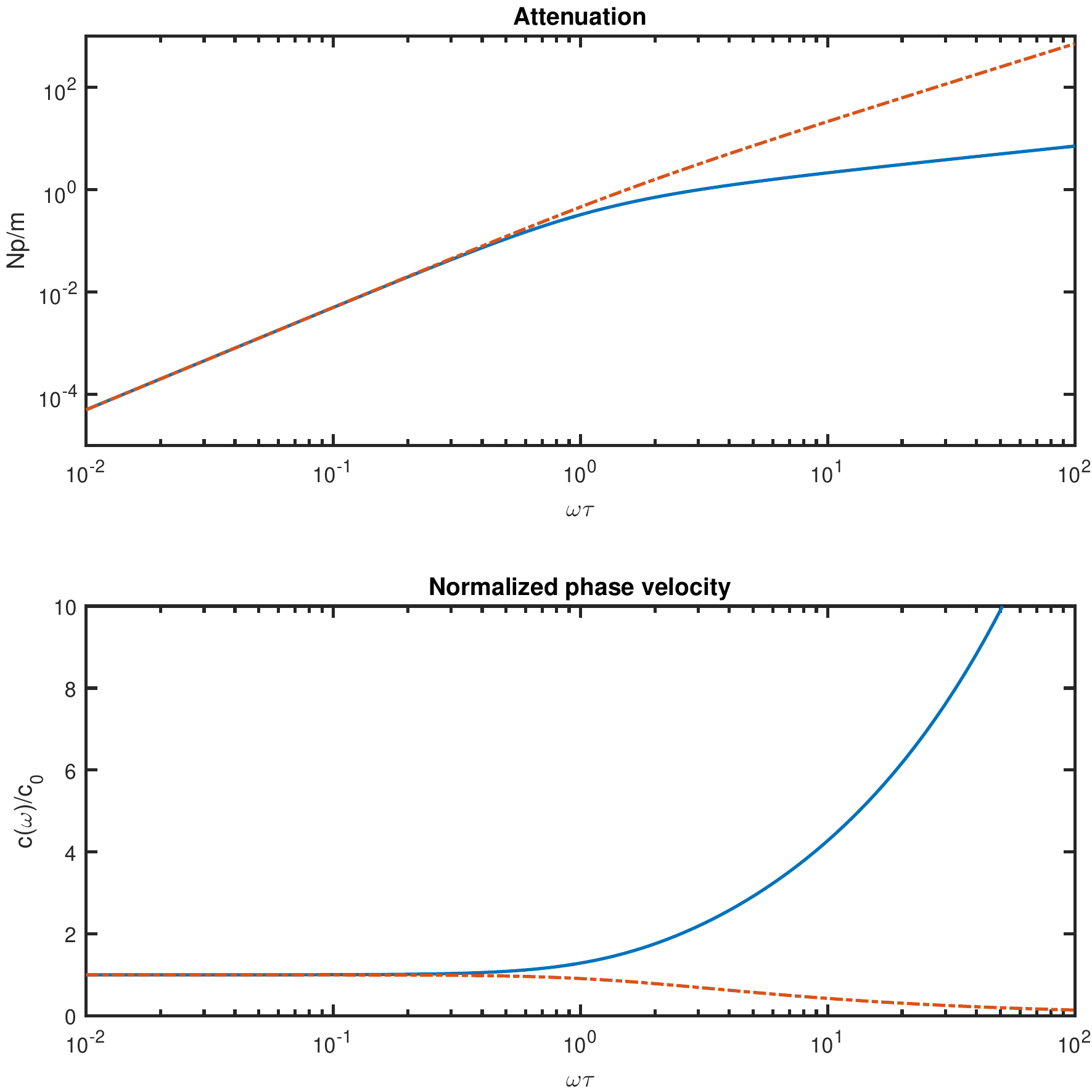}
  \caption{Viscous wave equation with mixed derivative loss term (solid line) compared to one with only temporal derivatives in the loss term (dash-dot line): Upper curve: Attenuation.  Lower curve: Phase velocity, both as a function of $\omega \tau$,  the normalized frequency (color online)}
  \label{fig:ApproxViscous}
\end{figure}
The dispersion equation is found by assuming a harmonic, plane wave solution in 1-D, $u(x,t)=\exp(i(\omega t - kx))$.
For the upper line of \eqref{eq:viscousX2} this gives:
\begin{equation}
k^2 - \frac{\omega^2}{c_0^2} + i \frac{\tau}{c_0^2} \omega^3 = 0 \quad \Rightarrow \quad k = \frac{\omega}{c_0} (1- i \omega \tau)^{1/2}.
\end{equation}
Assuming low frequencies/low losses, $\omega \tau \ll 1$, the dispersion equation to the second order approximation is:
\begin{equation}
k \approx \frac{\omega}{c_0} \left(1-i\frac{\omega\tau}{2} + \frac{(\omega\tau)^2}{8} \right).
\label{eq:lo-dispersion}
\end{equation}
Since the wave number can be expressed as in the first part of \eqref{eq:knumberDecomposed}, this gives
\begin{equation}
\alpha_{low}(\omega) = \frac{\tau}{2c_0} \omega^{2}, \quad c_{p,low}(\omega) = c_0(1-\frac{(\omega\tau)^2}{8}).
\label{eq:low-alpha}
\end{equation}
For high frequencies, $\omega \tau \gg 1$, the wavenumber is approximately 
\begin{align}
k \approx \frac{\omega}{c_0} \left(- i \omega \tau \right)^{1/2}
= \omega^{3/2} {\tau}^{1/2} \left[ \cos\left(\frac{\pi}{4}\right)  - i  \sin\left(\frac{\pi}{4}\right)  \right].
\end{align}
which results in:
\begin{equation}
\alpha_{high}(\omega) = \tau^{1/2} \sin\left(\frac{\pi}{4}\right) \omega^{3/2}, \quad 
c_{p,high}(\omega)= \frac{\tau^{-1/2} }{\cos\left(\frac{\pi}{4}\right)} \omega^{-1/2}.
\label{eq:high-alpha}
\end{equation}
Thus the attenuation increases proportionally with $\omega^2$ for low frequencies and the phase velocity decreases. For the high frequency/high loss case, the attenuation increases in proportion to $\omega^{3/2}$ and the phase velocity  decreases with frequency. The properties are illustrated in Fig.~\ref{fig:ApproxViscous} with the dash-dot line. These results violate the conditions of both \eqref{eq:asymptoteAttenuation} and \eqref{eq:asympoteDispersion} and implies that the upper version of the wave equation of \eqref{eq:viscousX2} is not derived from a spring damper constitutive equation and is  probably not passive.


\subsection{Mixed derivatives in loss term}

For the wave equation of the lower line of \eqref{eq:viscousX2} the dispersion equation is:
\begin{equation}
k^2 - \frac{\omega^2}{c_0^2} + i\omega\tau k^2 = 0 \quad \Rightarrow \quad k = \frac{\omega}{c_0} \frac{1}{(1+i \omega\tau)^{1/2}}.
\label{eq:dispersionViscous}
\end{equation}
For low frequencies/low losses, $\omega \tau \ll 1$,  the approximate wavenumber is 
\begin{equation}
k \approx \frac{\omega}{c_0} \left(1-i\frac{\omega\tau}{2} - \frac{3(\omega\tau)^2}{8} \right).
\label{eq:lo-dispersion2}
\end{equation}
The asymptotic value for attenuation is the same as for the previous case as expected, but the phase velocity now increases with frequency:
\begin{equation}
c_{p,low}(\omega) = c_0(1+\frac{3}{8}(\omega\tau)^2).
\end{equation}
For high frequencies/high losses the approximate wave number differs from the previous case
\begin{equation}
 k \approx \frac{\omega}{c_0} (i \omega\tau)^{-1/2} = 
		   \frac{\omega \tau^{-1/2}}{c_0} \left( \cos\frac{\pi}{4} -i \sin\frac{\pi}{4}\right) \omega^{-1/2}.
		   \label{eq:hi-dispersion}
\end{equation}
which results in
\begin{equation}
\alpha_{high}(\omega) = \frac{\sqrt{2}\tau}{2 c_0} \omega^{1/2} \quad c_{p,high}(\omega)= 2 c_0\sqrt{\frac{\tau}{2}} \omega^{1/2}.
\label{eq:high-alpha2}
\end{equation}

Thus the attenuation and the phase velocity both  increase asymptotically with $\sqrt{\omega}$ above a crossover frequency. Also the phase velocity is a non-decreasing function of frequency. These characteristics are seen in the solid line plots of Fig.~\ref{fig:ApproxViscous} and are in agreement with the conditions of  both \eqref{eq:asymptoteAttenuation} and \eqref{eq:asympoteDispersion}.


It should not be surprising that the  wave equation with a mixed derivative viscous term is the more physical one. 
%
It can be derived from a constitutive equation consisting of a spring and a damper in parallel, the Kelvin-Voigt model of Fig.~\ref{fig:Kelvin-Voigt_Zener} (left-hand side). 
The dynamic modulus, which is the Fourier transform of the impulse response of \eqref{eq:convH(t)Physical}, is $H(\omega) = \sigma(\omega)/\epsilon(\omega) = E + i \eta \omega$. $G(\omega)$ derives from a step response given by \eqref{eq:Constitutive} and when inserted into \eqref{eq:dispersionPhysical} the result is \eqref{eq:dispersionViscous} with $c_0^2 = E/\rho_0$, 
showing that this viscous wave equation indeed builds on a physical constitutive equation.

\section{Discussion}

The example in the previous section illustrates the usefulness of the asymptotic result.  The first analysis, that of \eqref{eq:viscousX2} with only temporal derivatives in the loss term, leads to the conclusion that there is no  constitutive equation of the spring damper type for it. The second analysis of the viscous wave equation does not give so definite an answer, due to the way the implications flow in Appendix \ref{app:Proof}. We cannot say for sure that there exists such a constitutive equation, as there is a possibility that the conditions of \eqref{eq:asymptoteAttenuation} and \eqref{eq:asympoteDispersion} may be satisfied in other ways also. In the case of the viscous wave equation, we have other ways of verifying that it indeed is rooted in a spring damper model, because it is straightforward to derive it from the Kelvin-Voigt model. Thus the lack of agreement with the asymptotic result tells us definitely that there is no spring-damper constitutive equation, but the opposite is more ambiguous \cite{Hanyga2013Wave}.

The spring damper constitutive equations all satisfy both  passivity criteria: positive dissipation and the more stringent Clausius-Duhem criterion. There is a possibility that there may be other medium models 
that are passive, and this is a topic which may be further investigated. But the spring damper or completely monotone class is the model behind two of the most common acoustic attenuation descriptions, the viscous loss and the relaxation loss models. It must therefore be the most important subclass of  passive systems.



One kind of medium where the requirement for complete monotonicity may be too strong is a porous medium. The Biot model predicts three wave modes: one shear mode, and two compressional modes. It has recently been shown that the shear wave solution is exactly equivalent to a Zener model \cite{holm2017spring}. However, the two compressional wave modes are only approximately equivalent to spring damper models. The solution to the fast wave equation is approximately that of a spring  in parallel with a Maxwell element, i.e.~a Zener model, and the slow wave can be approximated to that of a Maxwell element \cite{geertsma1961some}. Per mode, these solutions therefore cannot be described exactly by two or three term spring-damper models and  it is an open question if they can be described by a more complex network of spring-damper models according to \eqref{eq:RelaxationSpectrum} or not. The passivity criterion in this case is also more complex to formulate as one of the features of the Biot model is that energy is converted from one compressional wave mode to the other as discussed in Ch.~3 of Ref.~\citenum{chotiros2017acoustics}.

Fractional wave equations, where the loss term is described by non-integer derivatives, are also of interest. As mentioned in Sec.~\ref{sec:CM} all the basic fractional constitutive equations are completely monotonous. Fractional wave equations which have been derived from e.g.~the fractional Newton, Kelvin-Voigt, or Zener constitutive equations therefore satisfy the asymptotic results of \eqref{eq:asymptoteAttenuation} and \eqref{eq:asympoteDispersion}. But some of the fractional wave equations which have been derived by  simply substituting non-integer derivatives for the integer order temporal or spatial derivatives in a conventional wave equation do not. This analysis has  already been performed for some of the proposed fractional wave equations \cite{holm2011, holm2013deriving, holm2014comparison}. 

An alternative to the poroelastic model for sediment acoustics is the grain shearing model \cite{buckingham2000wave}. It has been shown that the grain shearing mechanism is a non-Newtonian linearly increasing viscosity which may be approximated by a fractional derivative \cite{pandey2016linking}. From that it follows that the shear wave model builds on a fractional Newton constitutive equation, and the compressional wave solution corresponds to a fractional Kelvin-Voigt model \cite{pandey2016connecting}. Furthermore its variants such as the viscous grain shearing model \cite{buckingham2007pore} can also be derived from a combination of ordinary and fractional spring damper models \cite{pandey2016connectingVGS}, so all of these models satisfy complete monotonicity.

In light of our derivation, Weaver and Pao\cite{Weaver1981} can also be reinterpreted. Their medium model is different  in that it is not a strain stress response model, but rather a propagating wave from one place in the material to another. Then they require that the transfer function  satisfies $H(\omega) \ge 0$ for passivity. That resembles more the dissipation-free criterion than the entropy-rate criterion. They show, in a way which is  different from ours, that the complex wave number, $k(\omega)$, then is a Herglotz function, which is also called a  Nevanlinna-Pick function. The set of complete Bernstein functions is actually a subset of these functions, 
%
%
see Theorem 6.7 in Ref.~\citenum{schilling2012bernstein}. From this property follows the asymptotic result for the attenuation.


\section{Conclusion}

Passive, fading memory materials which satisfy the second law of thermodynamics  belong to  a subset of those materials that can be described by a linear differential equation. Passivity  is also a stricter criterion than causality for a linear system and implies that there are restrictions on the relaxation modulus of the material and its first two derivatives.  The viscous and relaxation models of acoustics, and most interesting  materials, satisfy an even stricter criterion with restrictions on all derivatives of the relaxation modulus. This is the class of completely monotone moduli where the models  consist of a system of springs and dampers. Furthermore, all the springs and dampers have positive parameters.

The important completely monotone  subclass of passive systems cannot give rise to arbitrary attenuation and dispersion. It is shown here that the attenuation as a function of frequency has to increase slower than a linear function. Likewise the phase velocity has to increase monotonically. The viscous wave equation was then analyzed. It was shown that if the loss term  only has temporal derivatives then there does not exist any physical constitutive equation for the material, while if the loss term has a mix of spatial and temporal derivatives it is compatible with passivity.



\section{Acknowledgement}

We thank Professor Alexander Lion for helpful discussions concerning passivity and the Clausius-Duhem inequality.

\appendix

\section{Mathematical properties}

\subsection{Definition and representation of a complete Bernstein function}\label{app:cbf}
A function $f:(0,\infty) \rightarrow R$ is a Bernstein function if $f(t) \geq 0$, all derivatives exist,  and the sign of the derivatives alternate as in \eqref{eq:BernsteinDerivatives}.
A Bernstein function has the following representation
\begin{equation}
f(t) = a + bt + \int_0^\infty (1-e^{tr})\hat{\mu}(dr)
\label{eq:CBFrepresentation}
\end{equation}
where $a,b \geq 0$ and $\hat{{\mu}}$ is a measure on $(0,\infty)$ satisfying
 $\int_{0}^{\infty}\min\{1,r\}\hat{\mu}(dr) < \infty$.

A Bernstein function is said to be complete if the Lévy measure $\hat{\mu}$ has a completely monotone density. A complete Bernstein function $f:(0,\infty) \rightarrow R$ has the following representation (Remark 6.4 in Ref.~\citenum{schilling2012bernstein})
\begin{equation}
f(t) = a + b t +t \int_0^\infty \frac{1}{t+r}\sigma(dr)
\label{eq:appBernsteinDecomp1}
\end{equation}
where $a, b \ge 0$, $a = f(0)$, $b=\lim_{t\to\infty} f(t)/t$ and $\sigma$ is a measure on $(0,\infty)$ such that $\int_0^\infty 1/(1+r) \sigma(dr) < \infty$.

\subsection{Proof of complete Bernstein  property of wavenumber}
\label{app:Proof}
	The proof of the  numbered points in Sec.~\ref{sec:BernsteinProperty} proceeds in these  steps \cite{seredynska2010relaxationdispersion}. 
	\begin{enumerate}
\item $G(t) \in CM \Rightarrow G(s) \in CM$. The Laplace transform $G(s)$ of a CM time domain function, $G(t)$, is also CM by Theorem 1.4 of Ref.~\citenum{schilling2012bernstein}.

\item $G(s) \in CM \Rightarrow sG(s) \in CBF$. If $G(s) \in CM$, then it is has a Stieltjes representation: $G(s) = \frac{a}{s} + b + \int_{0}^{\infty} \frac{1}{s+r}\mu(dr)$ where $a$, $b \geq 0$ are non-negative constants and $\mu$ is a measure on $(0,\infty)$ such that $\int_0^\infty 1/(1+r)\mu(dr) < \infty$, see Definition 2.1 in Ref.~\citenum{schilling2012bernstein}. It follows that $sG(s)= a + bs + \int_{0}^{\infty} \frac{s}{s+r}\mu(dr)$ is a complete Bernstein function since it is equivalent to \eqref{eq:appBernsteinDecomp1}. 
	
\item $sG(s) \in CBF \Rightarrow \sqrt{sG(s)} \in CBF$. Corollary 7.6 in Ref.~\citenum{schilling2012bernstein} states that if $f_1$, $f_2$ are CBF, then $f_1(f_2)$ is CBF. Since $s^\alpha$ with $0 \leq \alpha \leq 1$ is CBF, $\sqrt{sG(s)}$ is also CBF.
	
\item $\sqrt{sG(s)} \in CBF \Rightarrow  \frac{s}{\sqrt{s G(s)}} \in CBF$. Theorem 6.2 in Ref.~\citenum{schilling2012bernstein} states that $\sqrt{sG(s)}/s$ is a Stieltjes function if $\sqrt{sG(s)}$ is CBF. Furthermore, the inverse of a Stieltjes function is a CBF by Theorem 7.3 in Ref.~\citenum{schilling2012bernstein}. As a result, $s/\sqrt{sG(s)}$ is CBF.
\end{enumerate}


%
%


\begin{thebibliography}{33}%
\makeatletter
\providecommand \@ifxundefined [1]{%
 \@ifx{#1\undefined}
}%
\providecommand \@ifnum [1]{%
 \ifnum #1\expandafter \@firstoftwo
 \else \expandafter \@secondoftwo
 \fi
}%
\providecommand \@ifx [1]{%
 \ifx #1\expandafter \@firstoftwo
 \else \expandafter \@secondoftwo
 \fi
}%
\providecommand \natexlab [1]{#1}%
\providecommand \enquote  [1]{``#1''}%
\providecommand \bibnamefont  [1]{#1}%
\providecommand \bibfnamefont [1]{#1}%
\providecommand \citenamefont [1]{#1}%
\providecommand \href@noop [0]{\@secondoftwo}%
\providecommand \href [0]{\begingroup \@sanitize@url \@href}%
\providecommand \@href[1]{\@@startlink{#1}\@@href}%
\providecommand \@@href[1]{\endgroup#1\@@endlink}%
\providecommand \@sanitize@url [0]{\catcode `\\12\catcode `\$12\catcode
  `\&12\catcode `\#12\catcode `\^12\catcode `\_12\catcode `\%12\relax}%
\providecommand \@@startlink[1]{}%
\providecommand \@@endlink[0]{}%
\providecommand \url  [0]{\begingroup\@sanitize@url \@url }%
\providecommand \@url [1]{\endgroup\@href {#1}{\urlprefix }}%
\providecommand \urlprefix  [0]{URL }%
\providecommand \Eprint [0]{\href }%
\providecommand \doibase [0]{http://dx.doi.org/}%
\providecommand \selectlanguage [0]{\@gobble}%
\providecommand \bibinfo  [0]{\@secondoftwo}%
\providecommand \bibfield  [0]{\@secondoftwo}%
\providecommand \translation [1]{[#1]}%
\providecommand \BibitemOpen [0]{}%
\providecommand \bibitemStop [0]{}%
\providecommand \bibitemNoStop [0]{.\EOS\space}%
\providecommand \EOS [0]{\spacefactor3000\relax}%
\providecommand \BibitemShut  [1]{\csname bibitem#1\endcsname}%
\let\auto@bib@innerbib\@empty
\bibitem [{\citenamefont {Weaver}\ and\ \citenamefont
  {Pao}(1981)}]{Weaver1981}%
  \BibitemOpen
  \bibfield  {author} {\bibinfo {author} {\bibfnamefont {R.~L.}\ \bibnamefont
  {Weaver}}\ and\ \bibinfo {author} {\bibfnamefont {Y.~H.}\ \bibnamefont
  {Pao}},\ }\bibfield  {title} {\enquote {\bibinfo {title} {{Dispersion
  relations for linear wave propagation in homogeneous and inhomogeneous
  media}},}\ }\href@noop {} {\bibfield  {journal} {\bibinfo  {journal} {Journ.
  {M}ath. {P}hys}\ }\textbf {\bibinfo {volume} {22}},\ \bibinfo {pages}
  {1909--1918} (\bibinfo {year} {1981})}\BibitemShut {NoStop}%
\bibitem [{\citenamefont {Buckingham}(2015)}]{buckingham2015frequency}%
  \BibitemOpen
  \bibfield  {author} {\bibinfo {author} {\bibfnamefont {Michael~J}\
  \bibnamefont {Buckingham}},\ }\bibfield  {title} {\enquote {\bibinfo {title}
  {Frequency power-law attenuation and dispersion in marine sediments},}\
  }\href@noop {} {\bibfield  {journal} {\bibinfo  {journal} {J.\ Acoust.\ Soc.\
  Am.}\ }\textbf {\bibinfo {volume} {137}},\ \bibinfo {pages} {2281--2281}
  (\bibinfo {year} {2015})}\BibitemShut {NoStop}%
\bibitem [{\citenamefont {Seredy\'{n}ska}\ and\ \citenamefont
  {Hanyga}(2010)}]{seredynska2010relaxationdispersion}%
  \BibitemOpen
  \bibfield  {author} {\bibinfo {author} {\bibfnamefont {M.}~\bibnamefont
  {Seredy\'{n}ska}}\ and\ \bibinfo {author} {\bibfnamefont {Andrzej}\
  \bibnamefont {Hanyga}},\ }\bibfield  {title} {\enquote {\bibinfo {title}
  {Relaxation, dispersion, attenuation, and finite propagation speed in
  viscoelastic media},}\ }\href@noop {} {\bibfield  {journal} {\bibinfo
  {journal} {J. Math. Phys.}\ }\textbf {\bibinfo {volume} {51}},\ \bibinfo
  {pages} {092901} (\bibinfo {year} {2010})}\BibitemShut {NoStop}%
\bibitem [{\citenamefont {Hanyga}(2013)}]{Hanyga2013Wave}%
  \BibitemOpen
  \bibfield  {author} {\bibinfo {author} {\bibfnamefont {A}~\bibnamefont
  {Hanyga}},\ }\bibfield  {title} {\enquote {\bibinfo {title} {Wave propagation
  in linear viscoelastic media with completely monotonic relaxation moduli},}\
  }\href@noop {} {\bibfield  {journal} {\bibinfo  {journal} {Wave Motion}\
  }\textbf {\bibinfo {volume} {50}},\ \bibinfo {pages} {909--928} (\bibinfo
  {year} {2013})}\BibitemShut {NoStop}%
\bibitem [{\citenamefont {Hanyga}(2014)}]{hanyga2014dispersion}%
  \BibitemOpen
  \bibfield  {author} {\bibinfo {author} {\bibfnamefont {Andrzej}\ \bibnamefont
  {Hanyga}},\ }\bibfield  {title} {\enquote {\bibinfo {title} {Dispersion and
  attenuation for an acoustic wave equation consistent with viscoelasticity},}\
  }\href@noop {} {\bibfield  {journal} {\bibinfo  {journal} {J. Comp. Acoust.}\
  }\textbf {\bibinfo {volume} {22}},\ \bibinfo {pages} {1450006} (\bibinfo
  {year} {2014})}\BibitemShut {NoStop}%
\bibitem [{\citenamefont {Hamilton}\ and\ \citenamefont
  {Blackstock}(1998)}]{Hamilton98}%
  \BibitemOpen
  \bibfield  {author} {\bibinfo {author} {\bibfnamefont {M.~F.}\ \bibnamefont
  {Hamilton}}\ and\ \bibinfo {author} {\bibfnamefont {D.~T.}\ \bibnamefont
  {Blackstock}},\ }\href@noop {} {\emph {\bibinfo {title} {{Nonlinear
  Acoustics}}}}\ (\bibinfo  {publisher} {Academic {P}ress},\ \bibinfo {address}
  {New York and London},\ \bibinfo {year} {1998})\BibitemShut {NoStop}%
\bibitem [{\citenamefont {Stokes}(1845)}]{Stokes1845}%
  \BibitemOpen
  \bibfield  {author} {\bibinfo {author} {\bibfnamefont {G.~G.}\ \bibnamefont
  {Stokes}},\ }\bibfield  {title} {\enquote {\bibinfo {title} {On the theories
  of the internal friction of fluids in motion, and of the equilibrium and
  motion of elastic solids},}\ }\href@noop {} {\bibfield  {journal} {\bibinfo
  {journal} {Trans. Cambridge Philos. Soc.}\ }\textbf {\bibinfo {volume} {8}}
  (\bibinfo {year} {1845})}\BibitemShut {NoStop}%
\bibitem [{\citenamefont {Mainardi}(2010)}]{Mainardi2010}%
  \BibitemOpen
  \bibfield  {author} {\bibinfo {author} {\bibfnamefont {Francesco}\
  \bibnamefont {Mainardi}},\ }\enquote {\bibinfo {title} {Fractional calculus
  and waves in linear viscoelesticity: An introduction to mathematical
  models},}\ \ (\bibinfo  {publisher} {Imperial College Press},\ \bibinfo
  {address} {London, UK},\ \bibinfo {year} {2010})\ pp.\ \bibinfo {pages}
  {1--347}\BibitemShut {NoStop}%
\bibitem [{\citenamefont {Tschoegl}(1989)}]{tschoegl1989phenomenological}%
  \BibitemOpen
  \bibfield  {author} {\bibinfo {author} {\bibfnamefont {Nicholas~W}\
  \bibnamefont {Tschoegl}},\ }\href@noop {} {\emph {\bibinfo {title} {The
  phenomenological theory of linear viscoelastic behavior: {A}n
  introduction}}}\ (\bibinfo  {publisher} {Springer-Verlag Berlin},\ \bibinfo
  {year} {1989})\ \bibinfo {note} {reprinted in 2012}\BibitemShut {NoStop}%
\bibitem [{\citenamefont {Boltzmann}(1876)}]{boltzmann1876theorie}%
  \BibitemOpen
  \bibfield  {author} {\bibinfo {author} {\bibfnamefont {Ludwig}\ \bibnamefont
  {Boltzmann}},\ }\bibfield  {title} {\enquote {\bibinfo {title} {Zur theorie
  der elastischen nachwirkung ({O}n the theory of hereditary elastic
  effects)},}\ }\href@noop {} {\bibfield  {journal} {\bibinfo  {journal} {Ann.
  Phys. Chem}\ }\textbf {\bibinfo {volume} {Bd. 7}},\ \bibinfo {pages}
  {624--654} (\bibinfo {year} {1876})}\BibitemShut {NoStop}%
\bibitem [{\citenamefont {Markovitz}(1977)}]{markovitz1977boltzmann}%
  \BibitemOpen
  \bibfield  {author} {\bibinfo {author} {\bibfnamefont {Hershel}\ \bibnamefont
  {Markovitz}},\ }\bibfield  {title} {\enquote {\bibinfo {title} {Boltzmann and
  the beginnings of linear viscoelasticity},}\ }\href@noop {} {\bibfield
  {journal} {\bibinfo  {journal} {Trans. Soc. Rheol. (1957-1977)}\ }\textbf
  {\bibinfo {volume} {21}},\ \bibinfo {pages} {381--398} (\bibinfo {year}
  {1977})}\BibitemShut {NoStop}%
\bibitem [{\citenamefont {Akyildiz}\ \emph {et~al.}(1990)\citenamefont
  {Akyildiz}, \citenamefont {Jones},\ and\ \citenamefont
  {Walters}}]{akyildiz1990spring}%
  \BibitemOpen
  \bibfield  {author} {\bibinfo {author} {\bibfnamefont {F}~\bibnamefont
  {Akyildiz}}, \bibinfo {author} {\bibfnamefont {RS}~\bibnamefont {Jones}}, \
  and\ \bibinfo {author} {\bibfnamefont {K}~\bibnamefont {Walters}},\
  }\bibfield  {title} {\enquote {\bibinfo {title} {On the spring-dashpot
  representation of linear viscoelastic behaviour},}\ }\href@noop {} {\bibfield
   {journal} {\bibinfo  {journal} {Rheol acta}\ }\textbf {\bibinfo {volume}
  {29}},\ \bibinfo {pages} {482--484} (\bibinfo {year} {1990})}\BibitemShut
  {NoStop}%
\bibitem [{\citenamefont {Triverio}\ \emph {et~al.}(2007)\citenamefont
  {Triverio}, \citenamefont {Grivet-Talocia}, \citenamefont {Nakhla},
  \citenamefont {Canavero},\ and\ \citenamefont
  {Achar}}]{triverio2007stability}%
  \BibitemOpen
  \bibfield  {author} {\bibinfo {author} {\bibfnamefont {Piero}\ \bibnamefont
  {Triverio}}, \bibinfo {author} {\bibfnamefont {Stefano}\ \bibnamefont
  {Grivet-Talocia}}, \bibinfo {author} {\bibfnamefont {Michel~S}\ \bibnamefont
  {Nakhla}}, \bibinfo {author} {\bibfnamefont {Flavio~G}\ \bibnamefont
  {Canavero}}, \ and\ \bibinfo {author} {\bibfnamefont {Ramachandra}\
  \bibnamefont {Achar}},\ }\bibfield  {title} {\enquote {\bibinfo {title}
  {Stability, causality, and passivity in electrical interconnect models},}\
  }\href@noop {} {\bibfield  {journal} {\bibinfo  {journal} {IEEE Trans. Adv.
  Packag.}\ }\textbf {\bibinfo {volume} {30}},\ \bibinfo {pages} {795--808}
  (\bibinfo {year} {2007})}\BibitemShut {NoStop}%
\bibitem [{\citenamefont {Waters}\ \emph {et~al.}(2005)\citenamefont {Waters},
  \citenamefont {Mobley},\ and\ \citenamefont {Miller}}]{waters2005causality}%
  \BibitemOpen
  \bibfield  {author} {\bibinfo {author} {\bibfnamefont {Kendall~R}\
  \bibnamefont {Waters}}, \bibinfo {author} {\bibfnamefont {Joel}\ \bibnamefont
  {Mobley}}, \ and\ \bibinfo {author} {\bibfnamefont {James~G}\ \bibnamefont
  {Miller}},\ }\bibfield  {title} {\enquote {\bibinfo {title}
  {Causality-imposed ({K}ramers-{K}ronig) relationships between attenuation and
  dispersion},}\ }\href@noop {} {\bibfield  {journal} {\bibinfo  {journal}
  {IEEE Trans.\ Ultrason.\ Ferroelectr.,\ Freq.\ Control}\ }\textbf {\bibinfo
  {volume} {52}},\ \bibinfo {pages} {822--823} (\bibinfo {year}
  {2005})}\BibitemShut {NoStop}%
\bibitem [{\citenamefont {Lion}(1997)}]{lion1997thermodynamics}%
  \BibitemOpen
  \bibfield  {author} {\bibinfo {author} {\bibfnamefont {Alexander}\
  \bibnamefont {Lion}},\ }\bibfield  {title} {\enquote {\bibinfo {title} {On
  the thermodynamics of fractional damping elements},}\ }\href@noop {}
  {\bibfield  {journal} {\bibinfo  {journal} {Continuum Mech. Therm.}\ }\textbf
  {\bibinfo {volume} {9}},\ \bibinfo {pages} {83--96} (\bibinfo {year}
  {1997})}\BibitemShut {NoStop}%
\bibitem [{\citenamefont {Haupt}\ and\ \citenamefont
  {Lion}(2002)}]{haupt2002finite}%
  \BibitemOpen
  \bibfield  {author} {\bibinfo {author} {\bibfnamefont {Peter}\ \bibnamefont
  {Haupt}}\ and\ \bibinfo {author} {\bibfnamefont {Alexander}\ \bibnamefont
  {Lion}},\ }\bibfield  {title} {\enquote {\bibinfo {title} {On finite linear
  viscoelasticity of incompressible isotropic materials},}\ }\href@noop {}
  {\bibfield  {journal} {\bibinfo  {journal} {Acta Mech}\ }\textbf {\bibinfo
  {volume} {159}},\ \bibinfo {pages} {87--124} (\bibinfo {year}
  {2002})}\BibitemShut {NoStop}%
\bibitem [{\citenamefont {Blackstock}(2000)}]{blackstock2000fundamentals}%
  \BibitemOpen
  \bibfield  {author} {\bibinfo {author} {\bibfnamefont {David~T}\ \bibnamefont
  {Blackstock}},\ }\href@noop {} {\emph {\bibinfo {title} {Fundamentals of
  physical acoustics}}}\ (\bibinfo  {publisher} {John Wiley \& Sons},\ \bibinfo
  {year} {2000})\BibitemShut {NoStop}%
\bibitem [{\citenamefont {Landau}\ and\ \citenamefont
  {Lifshitz}(1976)}]{landau1976mechanics}%
  \BibitemOpen
  \bibfield  {author} {\bibinfo {author} {\bibfnamefont {Lev~Davidovich}\
  \bibnamefont {Landau}}\ and\ \bibinfo {author} {\bibfnamefont
  {Evgenii~Mikhailovich}\ \bibnamefont {Lifshitz}},\ }\href@noop {} {\emph
  {\bibinfo {title} {{Mechanics, 3rd Edition: Vol. 1 of Course of Theoretical
  Physics}}}}\ (\bibinfo  {publisher} {Elsevier},\ \bibinfo {year}
  {1976})\BibitemShut {NoStop}%
\bibitem [{\citenamefont {Holm}\ and\ \citenamefont
  {N\"asholm}(2011)}]{holm2011}%
  \BibitemOpen
  \bibfield  {author} {\bibinfo {author} {\bibfnamefont {Sverre}\ \bibnamefont
  {Holm}}\ and\ \bibinfo {author} {\bibfnamefont {Sven~Peter}\ \bibnamefont
  {N\"asholm}},\ }\bibfield  {title} {\enquote {\bibinfo {title} {A causal and
  fractional all-frequency wave equation for lossy media},}\ }\href@noop {}
  {\bibfield  {journal} {\bibinfo  {journal} {J.\ Acoust.\ Soc.\ Am.}\ }\textbf
  {\bibinfo {volume} {130}},\ \bibinfo {pages} {2195--2202} (\bibinfo {year}
  {2011})}\BibitemShut {NoStop}%
\bibitem [{\citenamefont {Schilling}\ \emph {et~al.}(2012)\citenamefont
  {Schilling}, \citenamefont {Song},\ and\ \citenamefont
  {Vondracek}}]{schilling2012bernstein}%
  \BibitemOpen
  \bibfield  {author} {\bibinfo {author} {\bibfnamefont {Ren{\'e}~L}\
  \bibnamefont {Schilling}}, \bibinfo {author} {\bibfnamefont {Renming}\
  \bibnamefont {Song}}, \ and\ \bibinfo {author} {\bibfnamefont {Zoran}\
  \bibnamefont {Vondracek}},\ }\href@noop {} {\emph {\bibinfo {title}
  {Bernstein functions: theory and applications}}}\ (\bibinfo  {publisher}
  {Walter de Gruyter},\ \bibinfo {year} {2012})\BibitemShut {NoStop}%
\bibitem [{\citenamefont {Gross}(1947)}]{gross1947creep}%
  \BibitemOpen
  \bibfield  {author} {\bibinfo {author} {\bibfnamefont {B}~\bibnamefont
  {Gross}},\ }\bibfield  {title} {\enquote {\bibinfo {title} {On creep and
  relaxation},}\ }\href@noop {} {\bibfield  {journal} {\bibinfo  {journal} {J
  Appl Phys}\ }\textbf {\bibinfo {volume} {18}},\ \bibinfo {pages} {212--221}
  (\bibinfo {year} {1947})}\BibitemShut {NoStop}%
\bibitem [{\citenamefont {Caputo}\ and\ \citenamefont
  {Mainardi}(1971)}]{caputo1971linear}%
  \BibitemOpen
  \bibfield  {author} {\bibinfo {author} {\bibfnamefont {M}~\bibnamefont
  {Caputo}}\ and\ \bibinfo {author} {\bibfnamefont {F}~\bibnamefont
  {Mainardi}},\ }\bibfield  {title} {\enquote {\bibinfo {title} {Linear models
  of dissipation in anelastic solids},}\ }\href@noop {} {\bibfield  {journal}
  {\bibinfo  {journal} {La Rivista del Nuovo Cimento (1971-1977)}\ }\textbf
  {\bibinfo {volume} {1}},\ \bibinfo {pages} {161--198} (\bibinfo {year}
  {1971})}\BibitemShut {NoStop}%
\bibitem [{\citenamefont {Holm}(2017)}]{holm2017spring}%
  \BibitemOpen
  \bibfield  {author} {\bibinfo {author} {\bibfnamefont {Sverre}\ \bibnamefont
  {Holm}},\ }\bibfield  {title} {\enquote {\bibinfo {title} {Spring-damper
  equivalents of the fractional, poroelastic, and poroviscoelastic models for
  elastography},}\ }\href@noop {} {\bibfield  {journal} {\bibinfo  {journal}
  {Submitted for publication, preprint arXiv:1703.09515}\ } (\bibinfo {year}
  {2017})}\BibitemShut {NoStop}%
\bibitem [{\citenamefont {Holm}\ and\ \citenamefont {Sinkus}(2010)}]{Holm2010}%
  \BibitemOpen
  \bibfield  {author} {\bibinfo {author} {\bibfnamefont {Sverre}\ \bibnamefont
  {Holm}}\ and\ \bibinfo {author} {\bibfnamefont {Ralph}\ \bibnamefont
  {Sinkus}},\ }\bibfield  {title} {\enquote {\bibinfo {title} {{A unifying
  fractional wave equation for compressional and shear waves}},}\ }\href@noop
  {} {\bibfield  {journal} {\bibinfo  {journal} {J.\ Acoust.\ Soc.\ Am.}\
  }\textbf {\bibinfo {volume} {127}},\ \bibinfo {pages} {542--548} (\bibinfo
  {year} {2010})}\BibitemShut {NoStop}%
\bibitem [{\citenamefont {Geertsma}\ and\ \citenamefont
  {Smit}(1961)}]{geertsma1961some}%
  \BibitemOpen
  \bibfield  {author} {\bibinfo {author} {\bibfnamefont {J}~\bibnamefont
  {Geertsma}}\ and\ \bibinfo {author} {\bibfnamefont {D.~C.}\ \bibnamefont
  {Smit}},\ }\bibfield  {title} {\enquote {\bibinfo {title} {Some aspects of
  elastic wave propagation in fluid-saturated porous solids},}\ }\href@noop {}
  {\bibfield  {journal} {\bibinfo  {journal} {Geophysics}\ }\textbf {\bibinfo
  {volume} {26}},\ \bibinfo {pages} {169--181} (\bibinfo {year}
  {1961})}\BibitemShut {NoStop}%
\bibitem [{\citenamefont {Chotiros}(2017)}]{chotiros2017acoustics}%
  \BibitemOpen
  \bibfield  {author} {\bibinfo {author} {\bibfnamefont {Nicholas~P}\
  \bibnamefont {Chotiros}},\ }\href@noop {} {\emph {\bibinfo {title} {Acoustics
  of the Seabed as a Poroelastic Medium}}}\ (\bibinfo  {publisher} {Springer},\
  \bibinfo {year} {2017})\BibitemShut {NoStop}%
\bibitem [{\citenamefont {Holm}\ \emph {et~al.}(2013)\citenamefont {Holm},
  \citenamefont {N{\"a}sholm}, \citenamefont {Prieur},\ and\ \citenamefont
  {Sinkus}}]{holm2013deriving}%
  \BibitemOpen
  \bibfield  {author} {\bibinfo {author} {\bibfnamefont {S.}~\bibnamefont
  {Holm}}, \bibinfo {author} {\bibfnamefont {S.~P.}\ \bibnamefont
  {N{\"a}sholm}}, \bibinfo {author} {\bibfnamefont {F.}~\bibnamefont {Prieur}},
  \ and\ \bibinfo {author} {\bibfnamefont {R.}~\bibnamefont {Sinkus}},\
  }\bibfield  {title} {\enquote {\bibinfo {title} {Deriving fractional acoustic
  wave equations from mechanical and thermal constitutive equations},}\
  }\href@noop {} {\bibfield  {journal} {\bibinfo  {journal} {Comput. Math.
  Appl.}\ }\textbf {\bibinfo {volume} {66}},\ \bibinfo {pages} {621--629}
  (\bibinfo {year} {2013})}\BibitemShut {NoStop}%
\bibitem [{\citenamefont {Holm}\ and\ \citenamefont
  {N{\"a}sholm}(2014)}]{holm2014comparison}%
  \BibitemOpen
  \bibfield  {author} {\bibinfo {author} {\bibfnamefont {Sverre}\ \bibnamefont
  {Holm}}\ and\ \bibinfo {author} {\bibfnamefont {Sven~Peter}\ \bibnamefont
  {N{\"a}sholm}},\ }\bibfield  {title} {\enquote {\bibinfo {title} {Comparison
  of fractional wave equations for power law attenuation in ultrasound and
  elastography},}\ }\href@noop {} {\bibfield  {journal} {\bibinfo  {journal}
  {Ultrasound. Med. Biol.}\ }\textbf {\bibinfo {volume} {40}},\ \bibinfo
  {pages} {695--703} (\bibinfo {year} {2014})}\BibitemShut {NoStop}%
\bibitem [{\citenamefont {Buckingham}(2000)}]{buckingham2000wave}%
  \BibitemOpen
  \bibfield  {author} {\bibinfo {author} {\bibfnamefont {Michael~J}\
  \bibnamefont {Buckingham}},\ }\bibfield  {title} {\enquote {\bibinfo {title}
  {Wave propagation, stress relaxation, and grain-to-grain shearing in
  saturated, unconsolidated marine sediments},}\ }\href@noop {} {\bibfield
  {journal} {\bibinfo  {journal} {J.\ Acoust.\ Soc.\ Am.}\ }\textbf {\bibinfo
  {volume} {108}},\ \bibinfo {pages} {2796--2815} (\bibinfo {year}
  {2000})}\BibitemShut {NoStop}%
\bibitem [{\citenamefont {Pandey}\ and\ \citenamefont
  {Holm}(2016{\natexlab{a}})}]{pandey2016linking}%
  \BibitemOpen
  \bibfield  {author} {\bibinfo {author} {\bibfnamefont {Vikash}\ \bibnamefont
  {Pandey}}\ and\ \bibinfo {author} {\bibfnamefont {Sverre}\ \bibnamefont
  {Holm}},\ }\bibfield  {title} {\enquote {\bibinfo {title} {Linking the
  fractional derivative and the {L}omnitz creep law to non-{N}ewtonian
  time-varying viscosity},}\ }\href@noop {} {\bibfield  {journal} {\bibinfo
  {journal} {Phys. Rev. E}\ }\textbf {\bibinfo {volume} {94}},\ \bibinfo
  {pages} {032606} (\bibinfo {year} {2016}{\natexlab{a}})}\BibitemShut
  {NoStop}%
\bibitem [{\citenamefont {Pandey}\ and\ \citenamefont
  {Holm}(2016{\natexlab{b}})}]{pandey2016connecting}%
  \BibitemOpen
  \bibfield  {author} {\bibinfo {author} {\bibfnamefont {Vikash}\ \bibnamefont
  {Pandey}}\ and\ \bibinfo {author} {\bibfnamefont {Sverre}\ \bibnamefont
  {Holm}},\ }\bibfield  {title} {\enquote {\bibinfo {title} {Connecting the
  grain-shearing mechanism of wave propagation in marine sediments to
  fractional order wave equations},}\ }\href@noop {} {\bibfield  {journal}
  {\bibinfo  {journal} {J.\ Acoust.\ Soc.\ Am.}\ }\textbf {\bibinfo {volume}
  {140}},\ \bibinfo {pages} {4225--4236} (\bibinfo {year}
  {2016}{\natexlab{b}})}\BibitemShut {NoStop}%
\bibitem [{\citenamefont {Buckingham}(2007)}]{buckingham2007pore}%
  \BibitemOpen
  \bibfield  {author} {\bibinfo {author} {\bibfnamefont {Michael~J}\
  \bibnamefont {Buckingham}},\ }\bibfield  {title} {\enquote {\bibinfo {title}
  {On pore-fluid viscosity and the wave properties of saturated granular
  materials including marine sediments},}\ }\href@noop {} {\bibfield  {journal}
  {\bibinfo  {journal} {J.\ Acoust.\ Soc.\ Am.}\ }\textbf {\bibinfo {volume}
  {122}},\ \bibinfo {pages} {1486--1501} (\bibinfo {year} {2007})}\BibitemShut
  {NoStop}%
\bibitem [{\citenamefont {Pandey}\ and\ \citenamefont
  {Holm}(2016{\natexlab{c}})}]{pandey2016connectingVGS}%
  \BibitemOpen
  \bibfield  {author} {\bibinfo {author} {\bibfnamefont {V}~\bibnamefont
  {Pandey}}\ and\ \bibinfo {author} {\bibfnamefont {S}~\bibnamefont {Holm}},\
  }\bibfield  {title} {\enquote {\bibinfo {title} {Connecting the viscous
  grain-shearing mechanism of wave propagation in marine sediments to
  fractional calculus},}\ }in\ \href@noop {} {\emph {\bibinfo {booktitle} {78th
  EAGE Conference and Exhibition 2016}}}\ (\bibinfo {year} {2016})\BibitemShut
  {NoStop}%
  \end{thebibliography}

\end{document}